\documentclass[useAMS,usenatbib,referee]{mn2e}
\usepackage{epsfig}

\newcommand{\etal}{et~al.}

\newcommand{\ionhy}{H{\sc ii}}

\newcommand{\water}{$\mbox{H}_{2}\mbox{O}$}
\newcommand{\methanol}{$\mbox{CH}_{3}\mbox{OH}$}

\newcommand{\degrees}{$^\circ$}
\newcommand{\kms}{$\mbox{km~s}^{-1}$}

\newcommand{\msol}{\mbox{M\hbox{$_\odot$}}}
\newcommand{\arcm}{$^{\prime}$}

\title[High resolution observations of 6.7-GHz methanol masers]
{High resolution observations of 6.7-GHz methanol masers with the LBA}
\author[Dodson, Ojha, Ellingsen]{R. Dodson$^{1,3}$, R. Ojha$^{2}$, S. P. 
Ellingsen$^{3}$\thanks{E-mail:rdodson@vsop.isas.jaxa.jp; rojha@atnf.csiro.au;
Simon.Ellingsen@utas.edu.au}\\
$^1$ ISAS, 3-1-1 Yoshinodai, Sagamihara, Japan\\
$^2$ ATNF, CSIRO, PO 76, Epping 1710, Australia\\
$^3$ School of Mathematics and Physics, University of Tasmania, 
     Private bag 37, Hobart, Tasmania 7000, Australia\\ }

\begin{document}



\maketitle


\begin{abstract}
  We have used the Australian Long Baseline Array (LBA) to produce
  milliarcsecond images of five sites of methanol (\methanol ) maser
  emission at 6.7~GHz. These are all sites that have linear
  morphologies at arcsecond resolutions, which have been hypothesised
  to be due to the masers forming in edge-on circumstellar disks. We
  find that a simple disk model cannot explain the observations. We
  discuss various alternatives, and suggest a new model which explains
  how linear velocity gradients can be produced in methanol masers
  that arise in planar shocks propagating nearly perpendicular to the
  line of sight.

\end{abstract}

\typeout{Ojha reference? Absolute velocity? Grandeous verbage?}


\section{Introduction}
\label{s:intro}

Massive stars play a crucial role in the evolution of galaxies. They
are the cauldrons where heavy elements are produced. They provide
turbulent energy to the interstellar medium, are involved in the
production and destruction of molecular clouds and may regulate the
rate of star formation. Despite their importance, the processes
leading to massive star formation are poorly understood
(e.g. \citet{garay_99}, \citet{mckee_03}). Apart from theoretical
complexities inherent in such a multiple-process problem
\citep{yorke_02}, the chief obstacle to our understanding is
observational -- massive star formation occurs in fewer regions (as
they form quickly) that are, on average, further away (than
regions of low-mass star formation) and are highly obscured at optical
wavelengths (\citet{churchwell_90}, \citet{plume_92}).

Interstellar masers of three molecules, \water , OH and methanol, are 
both powerful and widely associated with active high-mass star formation 
regions (\citet{forster_89}, \citet{menten_91}). Thus, they have long 
been recognized as a potentially powerful tool to study the physical 
conditions and kinematics of massive star formation sites. However, the 
complex distribution of \water\ and OH masers has limited their role in 
the elucidation of massive star formation. Further, both \water\ and OH 
masers are also found towards other objects such as evolved low-mass stars 
\citep{wilson_72} and the centres of galaxies \citep{santos_79}.

The two strongest class II methanol masers (i.e. those closely
associated with OH masers and strong far-infrared emission) were
discovered at 12.178593~GHz by \citet{batrla_87} and at 6.668518~GHz by
\citet{menten_91}. These methanol masers are closely associated with
high-mass star forming regions \citep{norris_88, walsh_97}.  Sensitive
searches for 6.7-GHz methanol masers towards low-mass star forming
regions \citep{minier_03},
and nearby active galaxies \citep{phillips_1998a,
darling_03} have failed to detect any. While untargeted
searches covering large areas of the Galactic Plane have detected many
new 6.7-GHz methanol masers not associated with known massive star
forming regions, to date none have been confirmed as being associated
with any other type of astrophysical object
\citep{caswell_96,ellingsen_96, szymczak_02}. Thus class II methanol
masers prove more efficacious for finding and studying high-mass star
forming regions than either \water\/ or OH masers.

The brightness of methanol masers allows high resolution
interferometric observations to be made, which provide accurate
measurements of position, velocity and dimensions of the individual
maser components. Previous interferometric observations have generally
used connected element interferometers \citep{norris_88, norris_93,
phillips_98, walsh_98} that offer resolutions of the order of
arcseconds (which can be improved upon by super resolving the velocity
features, see \citet{phillips_98}, hereafter P98), though some VLBI
(Very Long Baseline Interferometry) experiments \citep{menten_88,
menten_92, norris_98, minier_98, moscadelli_99} with milliarcsecond
resolution have been performed. VLBI resolution images are vital to
uncovering the structure of these masing regions and their paucity
comes from the absence of telescopes (let alone arrays of telescopes)
able to observe at these frequencies until relatively recently. This
situation has now changed with both the Australian Long Baseline Array
(LBA\footnote{The Long Baseline Array is part of the Australia
Telescope which is funded by the Commonwealth of Australia for
operation as a National Facility managed by CSIRO.}, \citet{ojha_03})
and the European VLBI Network (EVN) covering the 6.7-GHz transition,
while the LBA and the Very Long Baseline Array (VLBA) covers the 12.2-GHz
transition.

Many of the observed masing sites have individual masers located along
lines or arcs, often with a near-monotonic velocity gradient along the
line. It has been suggested \citep{norris_93, norris_98, phillips_98}
that these linear structures trace masers embedded in an edge-on disk
surrounding a young star. Values of disk radii and enclosed masses
derived from modelling these sources assuming Keplerian rotation,
agree with theoretical models of accretion disks around massive stars
\citep{lin_90, hollenbach_94}.

\citet{walsh_98} found 36 of 97 maser sites to be linearly
extended and, while the circumstellar disk hypothesis is consistent
with their data, they considered it unlikely as the derived values of
the enclosed mass are too high in some cases. They suggested the
linear geometry of 6.7-GHz methanol masers resulted from shocks
arising behind a smooth shock front. As shocks would not produce the
velocity gradient they saw in many of their maser sites,
\citet{norris_93} had rejected shocks as a possible explanation.
Pointing out that most maser sites in their larger sample do not show
a systematic velocity gradient, \citet{walsh_98} argue that a velocity
gradient is not typical in maser sites. The VLBI observations of
\citet{minier_00} find the linear extent of maser regions to be about
10 times smaller than that seen at lower resolutions. This yields
sub-solar masses for the ``massive'' star. Both edge-on shock and edge-on disk
models produce linear features across the sky but shock fronts are
not normally so well ordered. At milliarcsecond (mas) resolutions we
probe resolutions of the order of tens of AU, the perfect scale to
investigate these differences.

The case for disks, or the other models, will only become clearer
after a significant increase in the sample size of methanol masers
imaged at very high resolution. We present the first VLBI images of
five sources which have not previously been imaged at mas
resolution. This allows us to directly compare conclusions based on
the ATCA (Australia Telescope Compact Array) data to those reached
with more than two orders of magnitude better resolution.

In Section \S{\ref{s:obsdata}} we describe our observations and the 
data reduction path. Our results are presented in \S{\ref{s:results}} 
and their implications for the current models are
discussed in \S{\ref{s:discuss}}. \S{\ref{s:conclude}} sets out our 
conclusions.




\begin{table*}
 \centering
\begin{minipage}{150mm}

\caption{Observed Maser Sites. 
Absolute positions for the strongest features are from (1)
Phillips \etal (1998) (2) Caswell (1997) and (3) Caswell \etal (1995).
Relative positions are with respect to these absolute
positions. Diameters, where calculated, are based on the nearer
kinematic distances and the enclosed masses were calculated from a
linear least-squares fit in a velocity-major axis diagram under the
assumption that the masers arise in an edge-on disk.}
\label{tab:one}

  \begin{tabular}{@{}lclrrrrrrc@{}}
  \hline
       &     Right   &               &Mean LSR   &          &Peak zero {\em uv} &          &      &  \\
Source &Ascension&Declination&Velocity& Distance & Flux Density & Diameter & Mass &Position\\
       &  (J2000)    &   (J2000) &(km~s$^{-1}$)& (kpc)    &   (Jy)   &   (AU)   & $\msol$  &  Reference \\
 \hline
G327.120$+$0.511 &  15:47:32.71 &$-$53:52:38.5 & $-$87.3   & 5.5/8.8  &  27.5    &1900      &17&1 \\
G327.402$+$0.444 & 15:49:19.50 &$-$53:45:13.9  & $-$81.8   & 5.1/9.2   &  233     & 239      &0.9&1 \\
G327.402$+$0.444E& 15:49:19.72 &$-$53:45:14.6  &          &           &  1.5     &   -      & -\\
G327.402$+$0.445 & 15:49:19.33 &$-$53:45:14.4  &          &           &  18      &  1392    & 9.2\\
G328.237$-$0.548 & 15:57:58.27  &$-$53:59:23.1 & $-$43.8  &  3.0     &  195     & 463      &0.4&1  \\
G328.237$-$0.548W& 15:57:58.20  &$-$53:59:23.2 &         &           &  5.5     & 560      & 2.5 \\
G328.254$-$0.532 & 15:57:59.73  &$-$53:58:00.8 & $-37.5$  & 2.6      &  47       &  141    & 0.0\\
G328.808$+$0.633 &  15:55:48.50 &$-$52:43:06.64 & $-$45.1 & 3.1      & 148      & 1713     &0.4&2\\
G328.809$+$0.633 &  15:55:48.62 &$-$52:43:06.55 &         &           & 29.9     & -        &- & \\
G329.029$-$0.205 &  16:00:31.80 &$-$53:12:49.7 & $-$37.3  &  2.6      &  114     &  164     &1.0&3\\
G329.031$-$0.198& 16:00:30.33 &$-$53:12:27.4 & $-$45.6   & 3.1       &  5.7     &  612     &  4.2     &  \\
\hline
\end{tabular}
\end{minipage}
\end{table*}

\section{Observations and Data reduction} 
\label{s:obsdata}

\subsection{Observations}
\label{s:obs}

The LBA is equipped to observe at 6.7~GHz at five of its 
antennas: Parkes, ATCA, Mopra, Hobart and Ceduna. Thus, at this frequency, 
this array provides baselines up to $\sim 1700$ km, yielding a resolution 
of 5.3 mas. 

The LBA uses the S2 recording system \citep{cannon_97,wilson_95}. We
used a bandwidth of 4-MHz centred on 6670 MHz. 
The LBA correlator formed 1024 channels across this bandwidth,
a velocity resolution of 0.2~\kms\ per channel. Both circular
polarizations were recorded, but only the parallel hand products were formed
as we had not included polarisation calibrators in the schedule. 

These observations were made on December 21$^{st}$ and 22$^{nd}$ 1999 for a
total of 10.5 hours.
Each target was typically observed for a total of 1.8 hours with six
`cuts' on the source. We observed a calibrator source on average every
eighty minutes in order to determine the delays and bandpasses.




\subsection{Methods}
\label{s:method}

Post correlation we loaded the visibilities into NRAO's Astronomical
Image Processing System ({\small AIPS}, \citet{aips}), where
we followed the conventional path for spectral line VLBI calibration
\citep{diamond_spec_vlbi}. Delays and rates were determined from the
calibrator scans, and the rates only were calculated for the target
sources.  
The velocities were corrected to the local standard of rest using the
{\small AIPS} task {\small CVEL}. The autocorrelation peak velocity
was corrected to match those found in the position references.
Amplitude calibration was achieved through scaling the maser
autocorrelations, which were output from the correlator at the same
time as the cross products. 
Once gain calibration was complete we exported the data for final
selfcalibration  in the Caltech package {\small DIFMAP}
\citep{difmap}. The relative positions shown in the figures are with
respect to the absolute positions given in Table~\ref{tab:one}.
 
 
Model fitting was performed on each channel independently by selecting
the minimum number of model parameters with a peak flux density
greater than five times the residual RMS (which is about 10~mJy/beam
in line free channels). All those model fits which were weak and
isolated in frequency and position were rejected. Care was taken to
search for all components within the field of view (2.5\arcm). We discovered
several maser emission clusters which had not been detected previously
as they merged with stronger emission at the same velocity in the
lower resolution ATCA observations.

The models fitted by {\small DIFMAP} were processed with {\small
MSPLOT} which was written by C.
Phillips\footnote{http://www.atnf.csiro.au/people/Chris.Phillips/software.html}
for just this kind of analysis.  Error estimates were based on the
synthesised beam size over the signal to noise (defined as the flux
per beam over residual image RMS). By model-fitting we can also
calculate the equivalent flux density at zero {\em uv} spacing, which
can be directly compared with the autocorrelation spectrum.


For those sources that were near to the phase centre and therefore
well imaged, there is no significant difference between the equivalent
zero spacing flux density and the flux density observed in the
autocorrelation spectra (taken from a subset of the Parkes
observations). We can therefore be confident that we have not missed
any significant maser emission in these regions and there is no need
to assume the existence of halos as in some other sources
\citep{minier_02}.

If we assume that the methanol maser emission is coming from an
edge-on disk we can calculate the contained mass from the velocity
gradient along the major axis, and therefore an upper limit on the
mass of the central star \citep{norris_98,walsh_98}. The fact that these
parameters derived from the ATCA observations agreed well with the
predicted masses and dimensions (tens of \msol\ and thousands of AU,
\citet{lin_90}) provided compelling support for the disk model.
%
We have used the kinematic distance model of \citet{msb_rot} to
estimate the distances of the masers based upon their flux-density
weighted mean velocity averaged across the arcsecond scale cluster.
This reduces the effect of the internal motions within the larger star
forming region.


\section{Results}
\label{s:results}

We observed five 6.7-GHz methanol maser sites, and detected eleven
clusters of emission within these sites. These clusters, their
positions and the mass of the central star calculated from the
methanol maser velocity gradient are listed in Table~\ref{tab:one}.

In the figures we plot the maser emission sites and clusters relative
to the position reference, with the spots sizes proportional to the flux
density, and 1-$\sigma$ error bars.  In the plots of the spatial
distribution of the maser emission the colour of the points represent
the velocity of the emission and the origin of the axes corresponds to
the position given in Table~\ref{tab:one}.  For most masers the error
in the relative position is smaller than the size of the points and so
the error bars are not visible. The line of best fit to the maser
emission is used to determine the major axis of the cluster.

For those sites of maser emission for which we determined the major
axis we have also produced plots of the maser velocity versus the
offset along the major axis. In these plots the colour of each point
represents it's distance (in mas) from the line of best fit in the
corresponding spatial plots (i.e. its offset in the minor axis
direction). The line of best fit to the velocity-major axis plot can
be used to estimate the enclosed mass in an edge-on disk model (see
Table~\ref{tab:one}).

We describe our results for each of these regions in the following
sections.



\subsection{G327.12+0.5}
\label{s:g327.12}

This maser cluster has previously been described at arcsecond
resolution by P98 who found it to be slightly offset from, but
projected on, an \ionhy\ region. They found the maser components
distributed along a line with a constant velocity gradient and their
spectrum shows the classic Keplerian shape. If we combine the maser
emission in individual channels we obtain a spatial distribution
matching that of P98.

\begin{figure}
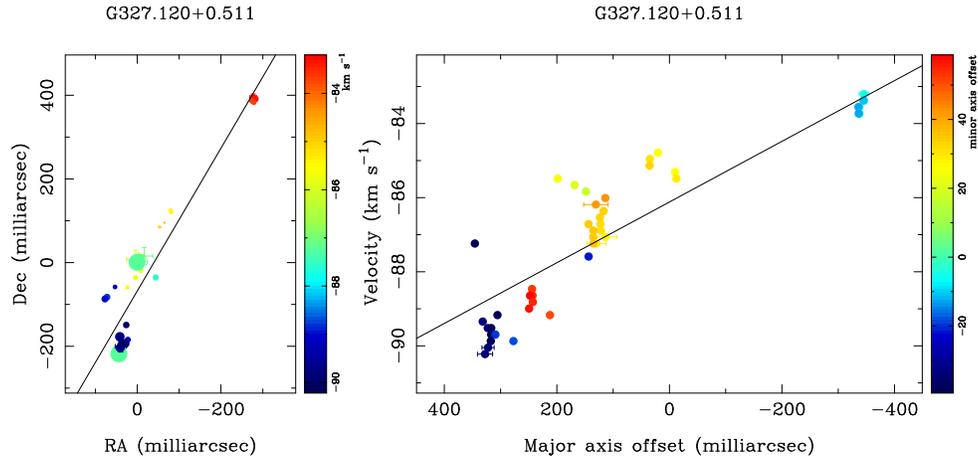

\begin{center}
\epsfig{file=g327.12+0.5.vlbi.ps, width=6cm, angle=270}
\epsfig{file=g327.12+0.5.va.ps, width=6cm, angle=270}
\caption{a) Maser positions for cluster G327.120+0.511 b)
Velocity-Major Axis plot of the same data. In a) we plot the maser emission
sites relative to the position reference with the spot sizes
proportional to flux density and 1-$\sigma$ error bars. The colours
represents the velocity of the spot. The line of best fit shows the
proposed major axis of a disk. In b) we show the velocity against the
distance along this major axis. The colour represents the distance
from this major axis. The line of best fit allows the determination of
the enclosed mass.}
\label{fig:g327.12}
\end{center}
\end{figure}

Our observations (Figure~\ref{fig:g327.12}) confirm the maser emission
distribution of P98, although the flux density has decreased
significantly. The peak flux density observed at the ATCA in 1994 was
73.6~Jy; we detect only 28~Jy. We find, however, that the concurrent
autocorrelation observations measure nearly the same levels as those
contained in our models. Therefore we are confident that source
variability is the cause of the difference.

\begin{figure}
\begin{center}
\epsfig{file=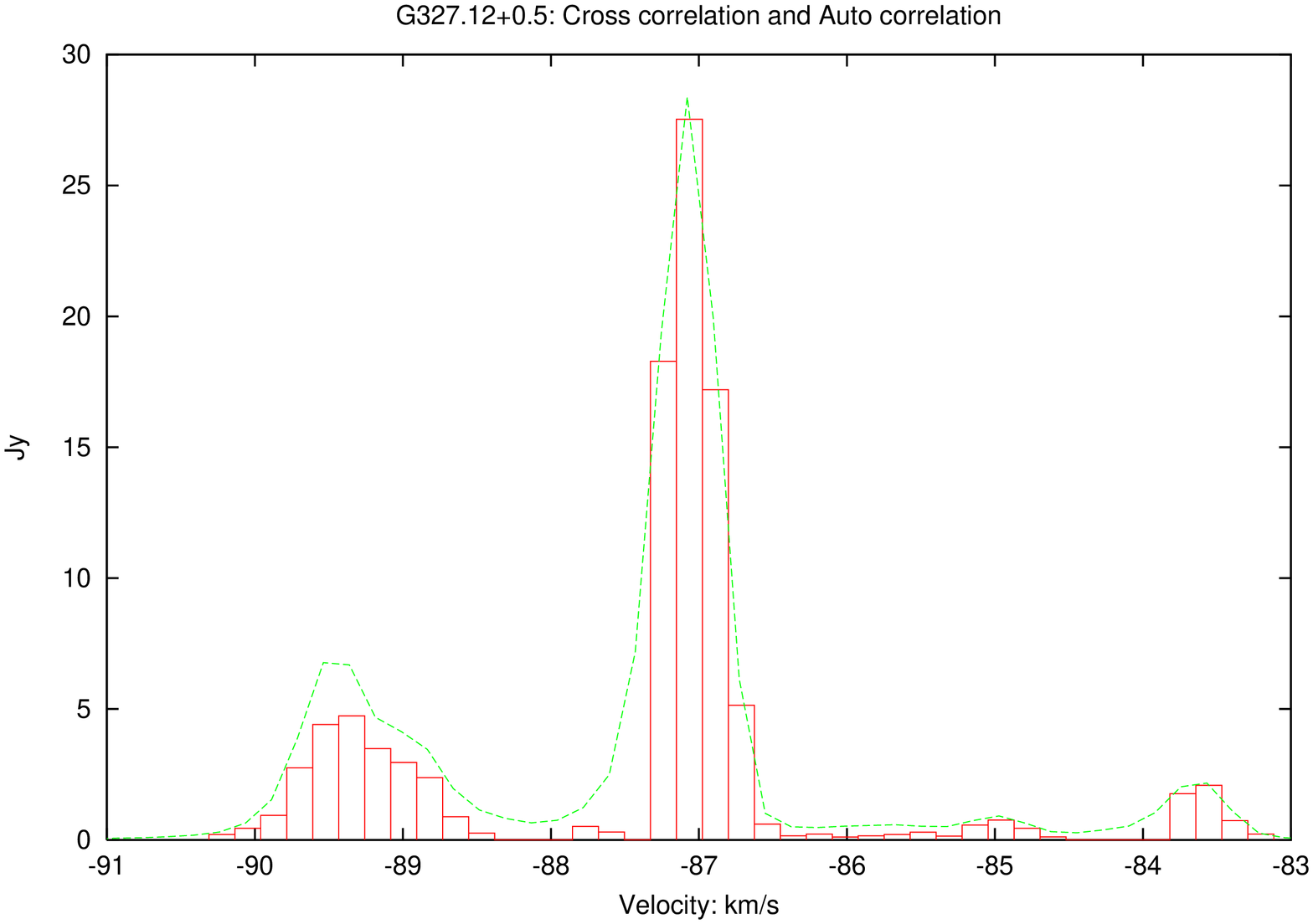, width=6cm, angle=270}
\caption{Autocorrelation (from Parkes with Antenna temperature
subtracted) (dotted line) against equivalent zero spacing flux density
(bar graph). This demonstrates the good flux recovery in the model
fitting.}
\label{fig:g327.12.ac}
\end{center}
\end{figure}

Our spectrum also shows the signature three-peaked profile of a
classic Keplerian disk (Figure~\ref{fig:g327.12.ac}), making this one
of the best candidates for the emission to define a disk.  We confirm
that the emission is from a single cluster, which extends across 690
mas with a RMS scatter of 21~mas from the line of best fit. The
equivalent size, at a kinematic distance of 5.5 or 8.8~kpc, is 1900~AU
or 3040~AU respectively. If gravitationally bound by a massive central
object, this implies an enclosed mass of 17.1 or 27.4 \msol\ . Whilst
the overall distribution lies along a straight line a closer
inspection of the individual components (see Figure
\ref{fig:g327.12}a) shows that the velocity structure is neither
smooth nor simple.  Another point of interest can be seen in the
velocity-major axis diagram (Figure~\ref{fig:g327.12}b) which shows
that the emission from individual channels tend to extend perpendicularly from
the major axis of the cluster.
We therefore
confirm that our VLBI observations are consistent with P98, but show
that the maser emission doesn't arise in a smooth disk and is very
disordered on mas scales.

%

\subsection{G327.40+0.4} 
\label{s:g327.40}

\begin{figure}
\begin{center}
\epsfig{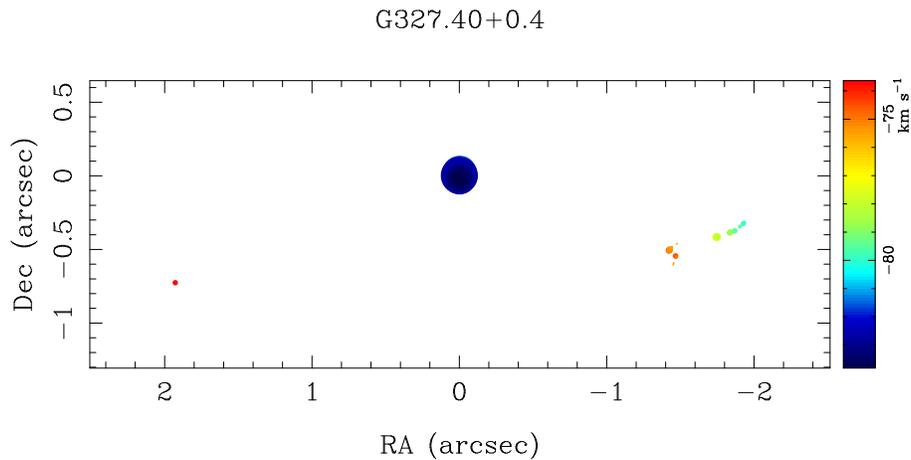}
\caption{Relative maser positions for cluster G327.40+0.44, using the
  same conventions as Figure~1a}
\label{fig:g327.40}
\end{center}
\end{figure}

P98 resolved this region into three separate clusters of emission;
G327.402+0.444, G327.402+0.445 and G327.402+0.444E. (Note that in P98
figure 10 the labels for G327.402+0.444 and G327.402+0.445 are
swapped.) We too detect maser emission from the same three sites.
P98 report G327.402+0.444E has only a single, weak, component, at
$-74$~\kms . The other two sites, G327.402+0.444 and G327.402+0.445,
have peak velocities of $-82.4$ and $-75.3$~\kms . Using the mean
velocity of the strongest maser cluster gives a near kinematic
distance of 5.1~kpc and a far distance of 9.2~kpc.  We have plotted
the relative positions of all the maser emission in Figure
\ref{fig:g327.40}.

\begin{figure}
\begin{center}
\epsfig{file=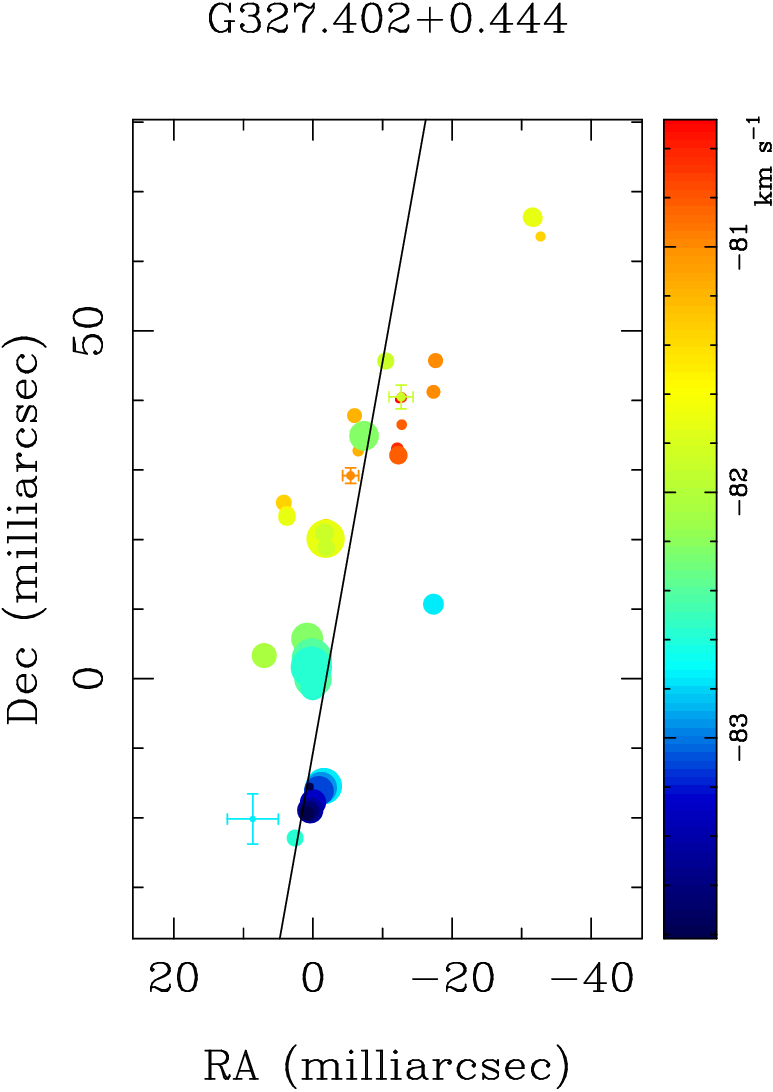, width=6cm, angle=270}
\epsfig{file=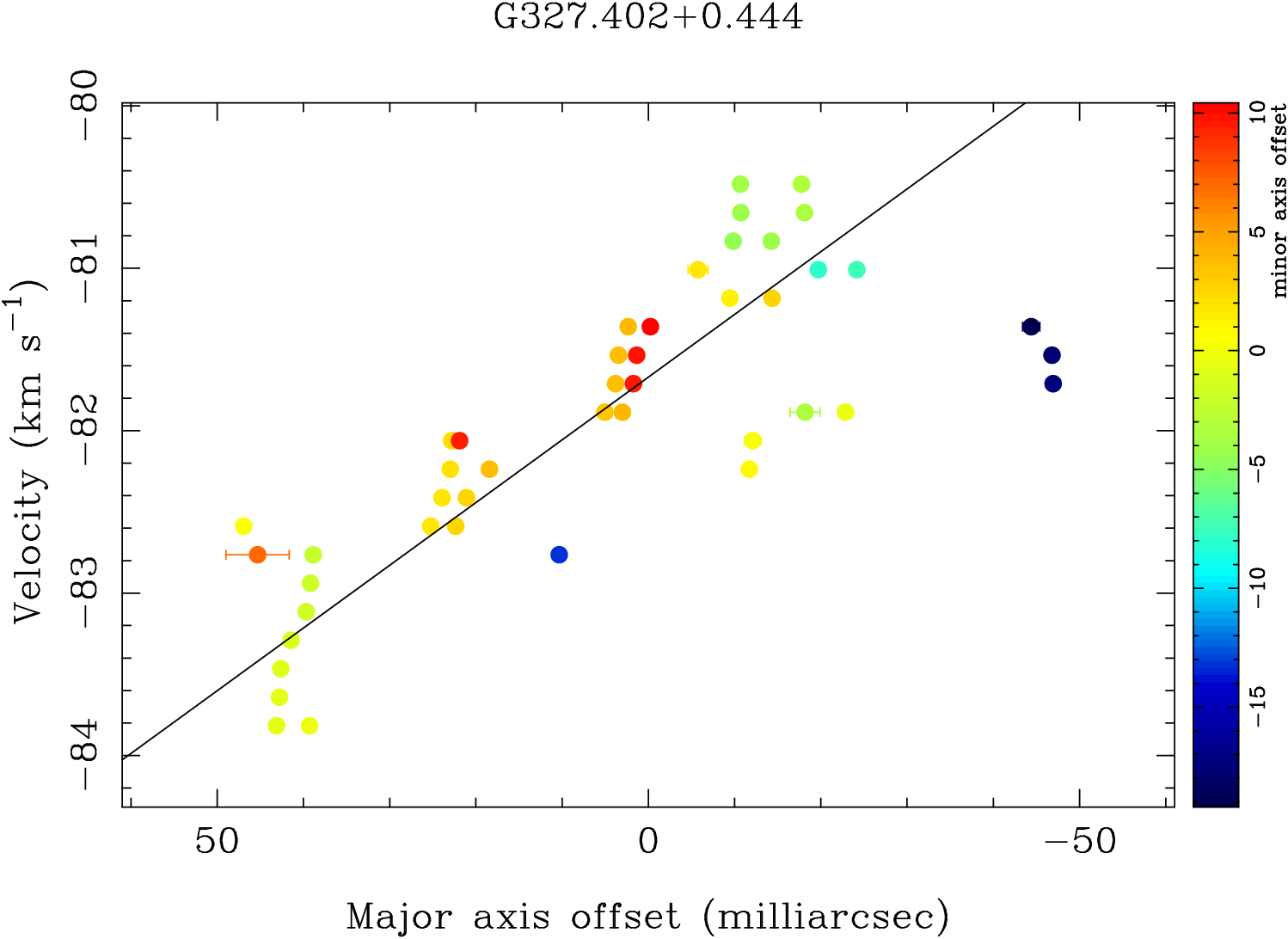, width=6cm, angle=270}
\caption{a) Maser positions for cluster G327.402+0.444 b) Velocity-Major Axis
plot of the same data. These follow the same conventions as Figure~1.}
\label{fig:g327.40+0.444}
\end{center}
\end{figure}

G327.402+0.444 (Figure~\ref{fig:g327.40+0.444}) exhibits a confused
morphology, with no clear velocity gradient along the major axis at a
position angle of 10\degrees . The velocity range is $-84$ to $-80.5$
\kms , and assuming the source is at the near kinematic distance of
5.1~kpc the linear size of the maser cluster is 239~AU.
G327.402+0.445 is a weaker cluster (peak flux density 230~Jy) with an
angular size of 540~mas at a position angle of 72 degrees (Figure
\ref{fig:g327.40+0.445}).
In P98 these two clusters are taken as separate sources, which we
agree with. They find that these sources both lie projected on an
\ionhy\ region with flux density of 199 mJy/beam.
For cluster G327.402+0.444 we did not detect the components labeled
A and B in P98, which alters the major axis considerably. The source
is known to be highly variable \citep{caswell_95}. 
As the axes for G327.402+0.444 and G327.402+0.445 are not aligned the
sources are not likely to be directly related and do not seem to form
the outer edges of a single body.  Furthermore, the enclosed mass for
the individual clusters is too small for them to be individual massive
star forming regions. The recovered flux density is shown in Figure
\ref{fig:g327.40.ac}.




\subsection{G328.24$-$0.5} 
\label{s:g328.24}

This region has been previously imaged by \citet{norris_93} and P98.
\citet{norris_93} report two sites, separated by 1 arcminute,
but overlapping in velocity. P98 report on both centres and describe
the Southern one as two ``clumps'' to the east and west of an \ionhy\ region
and with a velocity separation of almost 10~\kms\ . They suggest this
morphology and large velocity difference may result from shock fronts,
possibly in a bipolar outflow. 

We find the same maser emission centres: G328.254$-$0.532,
G328.237$-$0.548 and G328.237$-$0.548W. The northern one
(G328.254$-$0.532) was too far from the phase centre to prevent phase
wrapping within the integration time on the longest baselines, so
images of this cluster are not presented, but the flux densities
measured are included in our comparisons with the
autocorrelations. This cluster (with velocities between $-37$ and
$-39$~\kms and also at $-50$~\kms ) has poorer flux density recovery
due to the smearing within each integration. Figure~\ref{fig:g328.24}
shows the imaged maser emission regions and
Figure~\ref{fig:g328.24.ac} the autocorrelations.



G328.237$-$0.548 (Figures \ref{fig:g328.236}a \& b), covering $-42$ to
$-47$~\kms , is confused and made up of four regions of emission. The
peak flux density is 45~Jy at $-44.5$~\kms\ . The angular size of
G328.237-0.548 is 300~mas with a position angle is 66\degrees , which
corresponds to 460~AU at the near kinematic distance of 3~kpc. The
enclosed mass would be 0.4 \msol . Again the velocity gradient within
these spots (where a spot is the emission from within one synthesised
beam) does not lie along the major axis, instead lying across
it.

G328.237$-$0.548W, (Figures \ref{fig:g328.236}c \& d) covering $-31$ to $-37$
\kms , is linear and made up of two regions of emission with a
position angle of $9$\degrees . Using the kinematic distance of 3~kpc
the angular extent is 380 mas. The enclosed mass is 2.5 \msol\ and the
projected physical extent is 560~AU.

In P98 these two sites are assume to be related, as they lie either
side of the peak of an \ionhy\ region. We find no compelling
supporting evidence of this, as neither the position angles of the two
sources nor velocity spread within the spots align.


\subsection{G328.81+0.6} 
\label{s:g328.81}

This very widely spaced cluster was observed with the ATCA in 1992
\citep{norris_93, norris_98}. They find the masers to be distributed 
along a line but with no clear velocity gradient. 

We find this region to be made up of two clusters (Figures
\ref{fig:g328.81} and \ref{fig:g328.808}); G328.809+0.633 and
G328.808+0.633, as pointed out by \citet{norris_93}. The region does
not display a simple velocity distribution. Our relative positions
agree with those published (see Figure~\ref{fig:g328.81}). The
velocity-major axis plot shows no simple overall structure. Again the spread
of velocities lies perpendicular to the major axis. At a distance of
3.1~kpc the larger cluster, G328.808+0.633, which covers 1.1
arcseconds, has an extent of 1713~AU. Figure~\ref{fig:g328.81.ac}
shows the autocorrelations.
%

\subsection{G329.03$-$0.2}
\label{s:329.03}

This cluster (Figure~\ref{fig:g329.03}) does not seem to have been
observed previously at arcsecond or better resolution.  Three clusters
form a right angled triangle; G329.029$-$0.205, G329.029$-$0.205E,
G329.029$-$0.205S which spans 130 mas.
%
%
The velocity gradients within these spots do not lie along, but
across, the cluster major axis. The near kinematic distance
corresponding to the mean velocity is 2.6~kpc, and with an angular
extent of 130~mas the physical size is 164~AU.


There is additional weak emission between the velocities of $-41$~\kms
and $-50$~\kms. The peak position ($\alpha_{J2000} =$ 16:00:30.33
$\delta_{J2000} = -$53:12:27.4. 13.1, $-22.5$ arcseconds away from the
phase centre) matches the OH maser site G329.03$-$0.20
\citep{caswell_95} and there are three methanol emission regions in the
later. As this emission is significantly offset from the phase centre
rigorous analysis was not possible, additional observations centred
on this region are required for more thorough investigation. Figure
\ref{fig:g329.03.ac} shows the autocorrelations.



\section{Discussion}
\label{s:discuss}

\citet{norris_88} found that the 12.2-GHz methanol masers they imaged
often appeared to be distributed along a line or curve with some 
maser sites displaying a monotonic change in velocity along this line. 
Their observations of the 6.7-GHz methanol transition 
\citep{norris_93} confirmed these findings: 10 of 16 sources had  
components along a line or arc and half of these linear sources had a 
velocity gradient along the line. Various instrumental errors that 
could have created such features have been ruled out (not least by 
using different techniques) and there are too many sources with this 
linear geometry for the alignment to have arisen by chance. 

There are at least three possible ways a linear distribution of masers
could arise in a star forming region; (i) the masers may be embedded
in an edge-on, proto-planetary disk around a young star(s), (ii) they
may be slung along a collimated outflow or jet from the star(s), or
(iii) they may lie along a shock front between regions with different
physical conditions.

Bipolar outflows are common in star formation regions
\citep{shepherd_96}. However, such outflows are either broad and
uncollimated with low velocities (few \kms ) or are well collimated,
high-velocity flows. The observed linear geometry implies a high
degree of collimation but the line of sight velocity of the methanol
masers are offset by at most 5-10~\kms\ from the thermal molecular
line emission from the same region.  Also, the limited proper motion
studies that have been undertaken find velocities of only a few
kilometers per second \citep{moscadelli_02}. \citet{norris_93,
norris_98} point out that there is no proposed mechanism for producing
the high level of collimation implied in linear maser regions and, in
any case, the high-energy processes that might produce such jets would
probably dissociate the molecules of methanol. Also, the outflow model
does not offer any explanation for the uniform gradient seen in some
of the linear maser regions. It predicts that the orientation of the
line of masers should be radial to the central star or \ionhy\ region.

Both dense dust \citep{mueller_02} and molecular gas
\citep{beckwith_93} are abundant in star forming regions. In
particular, the region between the shock and ionization fronts around
an \ionhy\ region may possess conditions conducive to methanol maser
emission as has been suggested for OH masers \citep{elitzur_92}. Thus,
the linear geometry observed in maser regions could be the result of
these masers tracing the boundary between the differing physical
conditions on either side of a smooth shock front. \citet{norris_93,
norris_98} argue against this model pointing out that the maser spots
do not form a ring around the (spherically symmetric) \ionhy\ region
and there is no evidence for the existence of smooth and regular shock
fronts. Another weakness of this model is that it does not account for
the monotonic velocity gradient seen in many sources. Finally, many
methanol masers are not associated with detectable \ionhy\ regions
(P98; Walsh et al. 1998).

Star formation models \citep{lin_90, hollenbach_94} frequently invoke
disks around young stars. Such disks have indeed been observed in a
number of cases \citep{mundy_96, wilner_96} and are known to be rich
in warm molecular gas \citep{beckwith_93}. Thus the conditions for
maser emission may well exist. If these disks are observed edge-on a
linear distribution of masers would result. Disks seen at a small angle
from the edge-on orientation could account for the curved
distributions seen in some masers.  \citet{norris_93, norris_98}
favour this scenario and describe two mechanisms that would account
for most of these sources being seen edge-on rather than face-on. They
calculate that the column depth required for strong maser radiation
can only be achieved in the plane of the disk. Another possibility
they point out is that the maser radiation cannot escape in any
direction other than the plane of the disk because the disk slices
through the surrounding \ionhy\ region. Any maser emission
perpendicular to the disk is likely to be absorbed by the \ionhy\
region which has been shown \citep{hollenbach_94} to be optically
thick at centimeter wavelengths. Both these models are
illustrated in Figure~12 of \citet{norris_98}. Citing a lack of any
difference between their 6.7-GHz and 12.2-GHz data (the 12.2-GHz data
should be less affected), they rule out the second model.  A final
prediction of the disk model is that the masers in a source should lie
along a diameter of the \ionhy\ region with the continuum peak
(coincident with the star(s)) near the center of the row of masers.

In support of their disk model \citet{norris_93, norris_98} point out 
that as a uniformly rotating solid structure a disk would have a 
uniform velocity gradient. However, the motion is unlikely to be so 
simple and such a gradient will not always be present. In their sample, 
about half of the sources with a linear distribution of masers also 
show a velocity gradient along this line. Such a gradient is difficult 
to account for in the other two models. 
A common feature of the velocity-major axis diagrams we have
constructed is that the velocity spread in individual maser spots is
not along the major axis (the direction of the larger scale velocity
gradient) and this is contrary to expectations for a simple models of
maser emission in an edge-on disk. 
%
The velocity gradient within the spot (where a spot is the emission
from within one synthesised beam) is illustrated in Figures 1b, 4b,
5b, 9b, 9d, 11b and 13b. In all cases the velocity spread is not along
the major axis.


Modeling the maser sources with linear structure assuming Keplerian
motion, \citet{norris_98} find the velocities and positions of the
masers consistent with them being embedded in a rotating disk around a
star. Their model also yields disk radii of a few thousands of AU and
enclosed masses of a few tens of \msol\ which both agree with
theoretical models of disks around massive stars \citep{lin_90},
strongly supporting the disk model. They attribute the absence of the
double or triple-peaked spectral profile, characteristic of a
Keplerian disk \citep{ponomarev_94}, to small scale velocity and
density perturbations within the disk that reduce the ratio of
rotational to turbulent velocity.  Finally, they suggest that the
methanol regions that are complex i.e. do not show a linear geometry,
may result from confusion of two adjacent sources (which higher
resolution observations would clarify) or, complex regions may be
intrinsically different from linear ones.

P98 discuss a larger sample of 45 sources (of which they have imaged
33 with the ATCA) and report that 17 of them exhibit a linear or curved
geometry with 12 of these 17 also displaying a velocity gradient along
the line.  Of 25 sources they find to be associated with \ionhy\
regions, 18 either have their maser components aligned or slightly
offset with respect to the peak of the continuum emission. They do not
find the reduction in the fraction of linear sources with lower peak flux
density that was predicted by \citet{norris_98}. While they find the
linear sources consistent with the disk model, they find it difficult
to account for sources with complex morphology and suggest the latter
arise from different conditions, possibly a shock front.

Using the ATCA to observe an {\em IRAS}-based sample of 6.7-GHz
methanol masers detected at Parkes, \citet{walsh_98} were able to
obtain information on more than 150 regions with arcsecond resolution.
Of these, only 97 sources had enough maser spots to allow a meaningful
study of their geometry. Sixty one of these sources have no linear
structure, 27 have some and only 9 have a well-defined linear
structure. Assuming Keplerian motion, they estimate the central mass
that will result in the observed radial velocities and find (within
the large uncertainties inherent in their methods) that while they
cannot rule out the disk model, the derived central mass is often
unrealistically high. \citet{walsh_98} favour the shock hypothesis
with the maser spots being dense knots of gas that the passage of the
shock has compressed and accelerated. Maser spots require large column
lengths with velocity coherence along the line of sight which an near
edge-on shock can produce.  The velocities observed in the shock model
are the projection of the shock velocity onto our line of sight. If
the shock is expanding across a homogeneous medium even a smooth
velocity gradient could be produced. This model would result in
different maser spot geometries (including lines and curves) depending
on local conditions. The shock model does not, in general, produce
monotonic velocity gradients and this was a strong argument against it
\citep{norris_98}. \citet{walsh_98} point out that only 12 of 97 sites
in their sample show such a gradient and conclude that a linear
velocity gradient is not a regular property of maser sites.

\citet{minier_00} addressed this issue at VLBI resolution, observing
14 star forming regions at 6.7 and 12.2~GHz using the EVN and VLBA
networks, respectively. Ten sources show linear morphology and
(mostly) a smooth velocity gradient, in line with the disk model.  The
sample of Minier \etal\ includes the remarkable source NGC7538 which
contains a long (50 mas) thin continuous region of 6.7- and 12.2-GHz
maser emission with a linear velocity gradient \citep{minier_98}.  The
velocity gradients observed by \citet{minier_00} are similar to those
measured by P98 and \citet{norris_98}, but the linear extents seen by
\citet{minier_00} are about ten times smaller resulting in sub-solar
masses for the putative massive star. Thus higher resolution VLBI
images seem to cast doubt on the disk hypothesis though,
\citet{minier_00} suggest they could be only seeing a fraction of the
rotating disk, which will underestimate the central mass. The fact
that we find the same maser sites as the ATCA observations and recover
the majority of the autocorrelation flux density shows that we are not
resolving any significant proportion of the disk that is masing on
larger scales.

\citet{minier_02} also compare their zero equivalent flux density to the
autocorrelated flux. They find significant shortfall in brightness on
the VLBI baselines which they interpret as halos around fully
saturated masers.  We find some, but much less significant, excess
emission in the autocorrelation. This suggests that either the maser
spots the we are observing have much less flux density in the halo
than the sample of Minier \etal\ or that our shorter baselines make us
more sensitive to the extended emission.

\begin{table*}
  \caption{The spot size of selected maser features.  The linear size has
  been calculated assuming the near kinematic distance in each case.}
  \begin{tabular}{lrrrrc} \hline
    Source         & Velocity & Flux~~~~~~~  & Angular size & Linear size & 
      Brightness \\
                   & (\kms)   & Density (Jy) & (mas)        & (AU)        & 
      Temperature (K) \\  [2mm] \hline
    G327.120+0.511 & -83.4     & 1.0          & 6.5          & 36          & 
      9.4 $\times 10^8$ \\
                   & -87.1     & 15.0         & 42.0         & 231         &
      3.5 $\times 10^8$ \\
                   & -90.0     & 0.4          & 14.1         & 78          &
      8.0 $\times 10^7$ \\ \hline
    G327.402+0.445 &  -75.2  & 14.6         & 26.6         & 136         &
      8.2 $\times 10^8 $\\ 
    G327.402+0.444 &  -80.8     & 29.4         & 5.1          & 26          &
      4.5 $\times 10^{10}$ \\
                   & -82.4    & 162.0        & 2.1          & 10          &
      1.6 $\times 10^{12}$ \\ \hline
    G328.254$-$0.532 & -36.2    & 1.9          & 9.9          & 26          &
      3.7 $\times 10^8$ \\
    G328.237$-$0.548 & -44.6     & 182.0        & 5.6          & 15          &
      2.3 $\times 10^{11}$ \\ \hline
    G328.808+0.633 & -44.0    & 50.0         & 35.4         & 110         &
      1.6 $\times 10^9$ \\
                   & -44.5     & 148.0        & 4.8          & 15          &
      2.6 $\times 10^{11}$ \\ \hline
    G329.029$-$0.205 & -37.4     & 112.0        & 4.3          & 11          &
      2.4 $\times 10^{11}$ \\
                   & -40.4   & 4.3          & 6.6          & 17          &          3.9 $\times 10^9$ \\ \hline \hline
    \end{tabular}
    \label{tab:size}
\end{table*}

As stated earlier we determined the relative positions of the maser
emission features through model fitting the emission in each spectral
channel, using the minimum number of model parameters that produced
residuals less than five times the RMS noise.  Some emission is well
modelled with a point source, but much of the emission is resolved or
partially resolved.  For nearly all the resolved emission a circular
Gaussian provides an acceptable fit to the data.  Table~\ref{tab:size}
gives the observed flux density and the size of the circular Gaussian
for a representative sample of spots in all sources.  The brightness
temperatures of the maser emission ranges over 4 orders of magnitude,
with the brightest exceeding $10^{12}$~K.  The observed linear sizes and
brightness temperatures in our sample are similar to those observed by
Minier et al. (2002), however, we have been able to satisfactorily
model the maser emission with a single component rather than a
core/halo structure.

In contrast to the (single-source) VLBI observation reported in
\citet{norris_98} which confirmed the linear geometry revealed by ATCA
and the Parkes Tidbinbilla Interferometer data, and suggested ATCA
maps were sufficient to determine the broad structure of maser
sources, our LBA observations are producing more questions than they
answer. There is less sign of a simple linear structure at these
resolutions (\S{\ref{s:results}}), and the striking feature is the
presence of internal velocity gradients perpendicular to the cluster
major axis. Calculation of the enclosed mass using the methods of
\cite{walsh_98} yield enclosed masses of only a few \msol\ (see
Table~1) casting further doubt on the disk hypothesis.  Clearly a new
model which more consistently explains the growing wealth of
observational information is required.

\subsection{A new model for linear methanol masers}
One of the main arguments that has been put up against the shock model
is that the velocity range for the masers is low (in comparison to
that observed for water masers) and there have been no suggestions as
to how to achieve a highly linear morphology (in many cases with a
monotonic velocity gradient) in the absence of highly collimated, high
velocity outflows. We will now put forward a model that addresses
these objections, inspired by our observations of velocity gradients
within individual maser spots which are perpendicular to the major
axis of the maser spatial distribution. As has previously been
suggested by \citet{walsh_98} and others, our model is that class
II methanol masers arise behind low-speed ($<$ 10~\kms\/) planar
shocks.  This is consistent with the observation that class II
methanol masers typically have velocities within 5-10~\kms\ of the
thermal emission from the same regions and also the proper motion
results of \citet{moscadelli_02}, which find velocities in the range
1-7~\kms\ for the class II methanol masers in W3(OH). To produce a
linear spatial distribution only requires that the shock is
propagating nearly perpendicular to the line of
sight. Figure~\ref{fig:model} illustrates the shock front.
Interaction of the shock with density perturbations in the star
forming region will gradually disrupt its linearity and reduce it with
time.  In this model the strongest methanol masers will arise behind
shocks that are perpendicular to the line of sight as this is when the
path length is greatest.  A prediction we can make from this model is
that  as the inclination angle of the shock increases we would expect
a) the peak flux density of the masers to decrease, b) the average linewidth
of the masers to increase and c) the internal velocity gradient within a
maser spot to increase.
This is because as the inclination angle of the shock-front increases,
a) the coherence length is reduced, b) a greater range of the
star forming region's velocity is sampled and c) the individual channels
can be super-resolved along the shock front.
The majority of methanol maser sources do not
show linear morphologies and these arise in shocks which are seen at
angles more highly inclined to the line of sight.  For an inclined
shock the path length will be shorter and so the chances of having a
sufficiently long path with line of sight velocity coherence is
decreased.  The masers will arise at random locations and show an
overall complex morphology.  Further, we know that shocks are common
in the high-mass cluster mode of star formation and so the presence of
linear structures with a small number of offset maser spots can be
explained as being due to different shocks.  In this case the proper
motion of the offset maser features should be systematically
different.

An obvious question is why haven't internal velocity gradients
perpendicular to the large scale velocity gradient been noticed in
previous VLBI observations of methanol masers?  There are two possible
explanations for this, the first is that many of the sources for which
VLBI observations have been made are those with the highest peak flux
density.  In the model we have proposed the strongest masers are those
which are closest to being perpendicular to the line of sight and so those
where the effect is smallest.  The second possible explanation relates
to the way in which maser images are often presented.  In many cases,
maser emission in consecutive spectral channels which are separated by
less than a synthesized beam are represented as a single maser spot at
some average position.  In doing this any information on the internal
velocity gradient is lost, it is only when data for each individual
spectral channel is plotted on a velocity-major axis diagram that
the orthogonal orientation of the internal and large scale velocity
gradients stands out.  

The final remaining question is how to produce a monotonic velocity
gradient within a planar shock front.  One possibility is that it
arises when an edge-on planar shock propagates into a rotating dense
molecular clump or star forming core.  In general the shock will make
some angle to the rotation axis of the clump/core and this will result
in a linear gradient in the observed line of sight velocity.  This is
because the line of sight velocity at the tangential point of the
clump is simply $V_{los} = r \omega$ where $r$ is the distance between
the shock front and the rotation axis of the clump and $\omega$ is the
angular frequency of the clump rotation. This is illustrated in Figure
\ref{fig:model}. The typical linear dimensions of a methanol maser
region are observed to be approximately 0.03 pc
(Caswell 1997, P98),
while the typical dimensions for cores
in high-mass star forming regions is 0.1 pc \citep{jijina_99}.  The
dimensions and temperatures of these cores are in the right range to
produce methanol masers, although the density and methanol abundance
must be orders of magnitude greater in the post-shocked gas than
values averaged across the core in order for masing to occur
\citep{cragg_01,cragg_02}. \citet{moscadelli_02} have undertaken the only
proper motion study of methanol masers and model the observed
kinematics as an expanding conical outflow.  They propose that the
linear feature seen in the 6.7- and 12.2-GHz methanol masers and
6.035-MHz OH masers in W3(OH) arises along one edge of the cone as
that produces paths approximately parallel to the line of sight.  This
is clearly an alternative method of producing linear structures, which
is consistent with a shock model for the methanol masers, however, if
it were the most prevalent method then we might expect to see X or V
like structures in some sources as well, but these have not been
observed.

The model we are putting forward to explain our observations, and as a
general explanation for the spatial morphologies observed in methanol
masers, leads to a number of predictions which enable it to be tested
through comparison with observations.  Our model predicts the
following, with the last two being the most definitive tests.

\begin{enumerate}
\item The internal velocity gradients within a maser spot should be
  perpendicular to the major axis of the spatial extension and of a
  similar magnitude for all spots.
\item The percentage of linear class II sources should be a function
  of the peak luminosity. 
\item The linewidth of the masers should decrease with luminosity and
  the magnitude of the internal velocity gradient increase.

\item The proper motion of the maser spots should be perpendicular to
  the major axis (compared to being along the major axis for an
  edge-on disk).
\item The magnetic field near a shock front should be oriented
  parallel to the shock and hence parallel to the direction of
  elongation, while for a disk the magnetic field is expected to thread
  the disk and hence be perpendicular to the direction of elongation.
  Polarization observations of the methanol masers should be able to
  distinguish between these two models, although it may be necessary
  to go to VLBI resolutions to avoid confusion of the position angle
  due to multiple maser spots at arcsecond resolutions.
\end{enumerate}

Our sample has only seven well imaged extended sources, but we can
investigate whether these match our proposed tests. Tests ii), iv) and v)
require further observations, but test i) is fulfilled.
For test iii) when we compare the linewidth against luminosity we find
the slope is $-0.8\pm 0.2 \times 10^{-4}$~\kms /Jy~kpc$^2$. The
outliers are the two narrowest sources, G328.808+0.633, and
G327.120+0.511 (see Figure \ref{fig:modfit}a).
%
To estimate the velocity gradients across the source we have grouped
those data points close in position (within 5 mas) and velocity, and
measured the velocity-major axis gradient. Figure~\ref{fig:modfit}b
plots the luminosity against the median gradient (in mas/\kms ). 
This shows that the velocity gradients are similar within the
clusters, but any correlation with luminocity is hidden by the scatter
amongst the sources. A larger sample is required to investigate this
proposed relationship properly.

%

\begin{table*}
\begin{minipage}{140mm}
\caption{Maser sites and their parameters. Luminosity is given in
  Jy~kpc$^2$ without any attempt to correct for beaming or distance
  ambiguities. Line widths are the full detectable velocity range of
  the central (i.e. brightest) emission group. The final column is the median
  velocity gradient in each identifiable group and the RMS scatter for
  that group.}
\label{tab:two}
\begin{tabular}{@{}lclrrc@{}}
\hline
Source &Distance&Flux Density       &Luminosity&line width&Velocity
       gradient range\\
       & kpc    &   (Jy)            &Jy~kpc$^2$& \kms & mas/\kms \\
\hline
G327.120$+$0.511 & 5.5/8.8   & 27.5 & 832/2130 & 0.4 & $\phantom{-}1\pm3$\\
G327.402$+$0.444 & 5.1/9.2   &  233 &6060/19720& 0.5 & $-3\pm4$\\
G327.402$+$0.445 &           &  18  &468/1523  & 1.0 & $12\pm4$\\
G328.237$-$0.548 &   3.0     &  195 &  1760    & 0.7 & $\phantom{-}0\pm5$\\
G328.237$-$0.548W&           &  5.5 &   50     & 1.1 & $-1\pm5$\\
G328.808$+$0.633 &  3.1      & 148  &  1420    & 0.4 & $\phantom{-}2\pm11$\\
G329.029$-$0.205 &  2.6      &  114 &  770     & 1.1 & $\phantom{-}0\pm4$\\
\hline
\end{tabular}
\end{minipage}
\end{table*}

Despite the wide variety of geometries seen in methanol spot
distribution in masing regions, the observed size and velocity ranges
are relatively narrow suggesting a common mechanism, or at least that
there are few astrophysical environments that give rise to class~II
methanol masers.  VLBI resolution imaging, plausibility arguments
relating to the number of sources and other inconsistencies suggest
that the disk model is an unlikely scenario to explain the majority of
linear methanol maser sources.  There exists little support in the
literature for the outflow hypothesis and there is growing evidence
from observations of H$_{2}$ that some methanol masers are associated
with shocked gas \citep{lee_01, debuizer_03}. Also De Buizer \etal\ 
find that the distribution of H$_2$ emission is predominantly parallel with
the linear maser structures, which is consistent with our hypothesis, but
not with a disk model. The shock hypothesis is consistent
with a wide range of geometries that arise from differing local
conditions, including linear and curved geometries in a fraction of
sources.

\section{Conclusions}
\label{s:conclude}

We have observed five sites of methanol maser emission, and imaged
eleven clusters with milliarcsecond resolution. These high resolution
images argue strongly that the linear structures commonly observed are
more likely linear shock fronts than Keplerian disks and some of the
predictions of the former model are fulfilled. We are, however, not
able to provide conclusive evidence of this, and VLBI observations of
a large sample is required. We have proposed a further series of
experiments in which we will collect full polarisation data which
should determine the relationships between the maser spots.

\section*{Acknowledgments}

This research has made use of NASA's Astrophysics Data System Abstract 
Service and the SIMBAD database, operated at CDS. 
The LBA is operated as a National Facility, and managed by CSIRO and
the University of Tasmania.
RD and SE were funded by ARC postdoctoral awards.
We wish to thank Dr P. Edwards for his assistance.


\onecolumn

\begin{figure}
\begin{center}
\epsfig{file=g327.402+0.445.vlbi.ps, width=4cm, angle=270}
\epsfig{file=g327.402+0.445.va.ps, width=4cm, angle=270}
\caption{a) Maser positions for cluster G327.402+0.445 b) Velocity-Major Axis
plot of the same data. These follow the same conventions as Figure~1.}
\label{fig:g327.40+0.445}
\end{center}
\end{figure}

\begin{figure}
\begin{center}
\epsfig{file=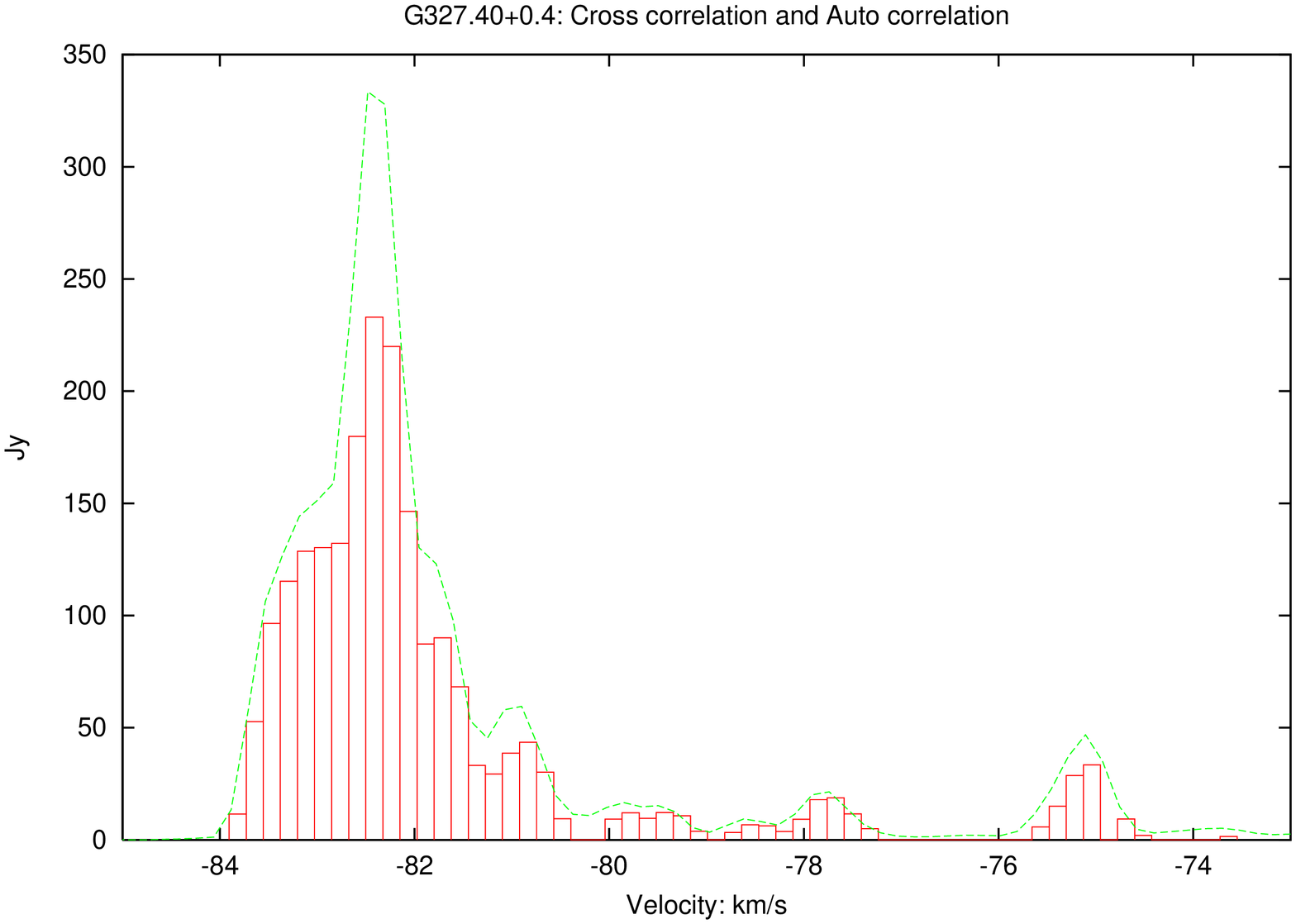, width=6cm, angle=270}
\caption{Autocorrelation (from Parkes with Antenna temperature
subtracted) (dotted line) against equivalent zero spacing flux density
(bar graph). This demonstrates
the good flux recovery in the model fitting.}
\label{fig:g327.40.ac}
\end{center}
\end{figure}

\begin{figure}
\begin{center}
\epsfig{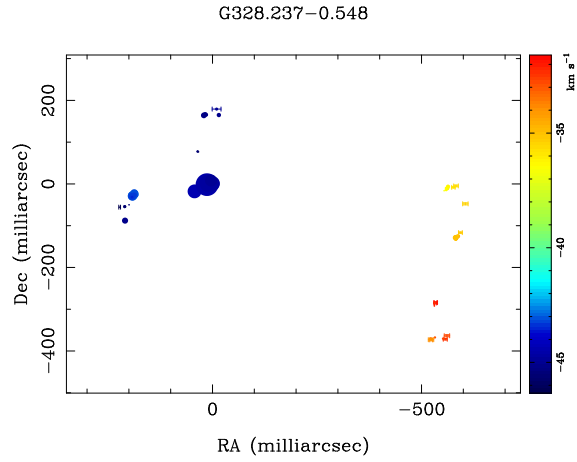}
\caption{Relative maser positions for cluster G328.237$-$0.548. These follow the same conventions as Figure~1.}
\label{fig:g328.24}
\end{center}
\end{figure}

\begin{figure}
\begin{center}
\epsfig{file=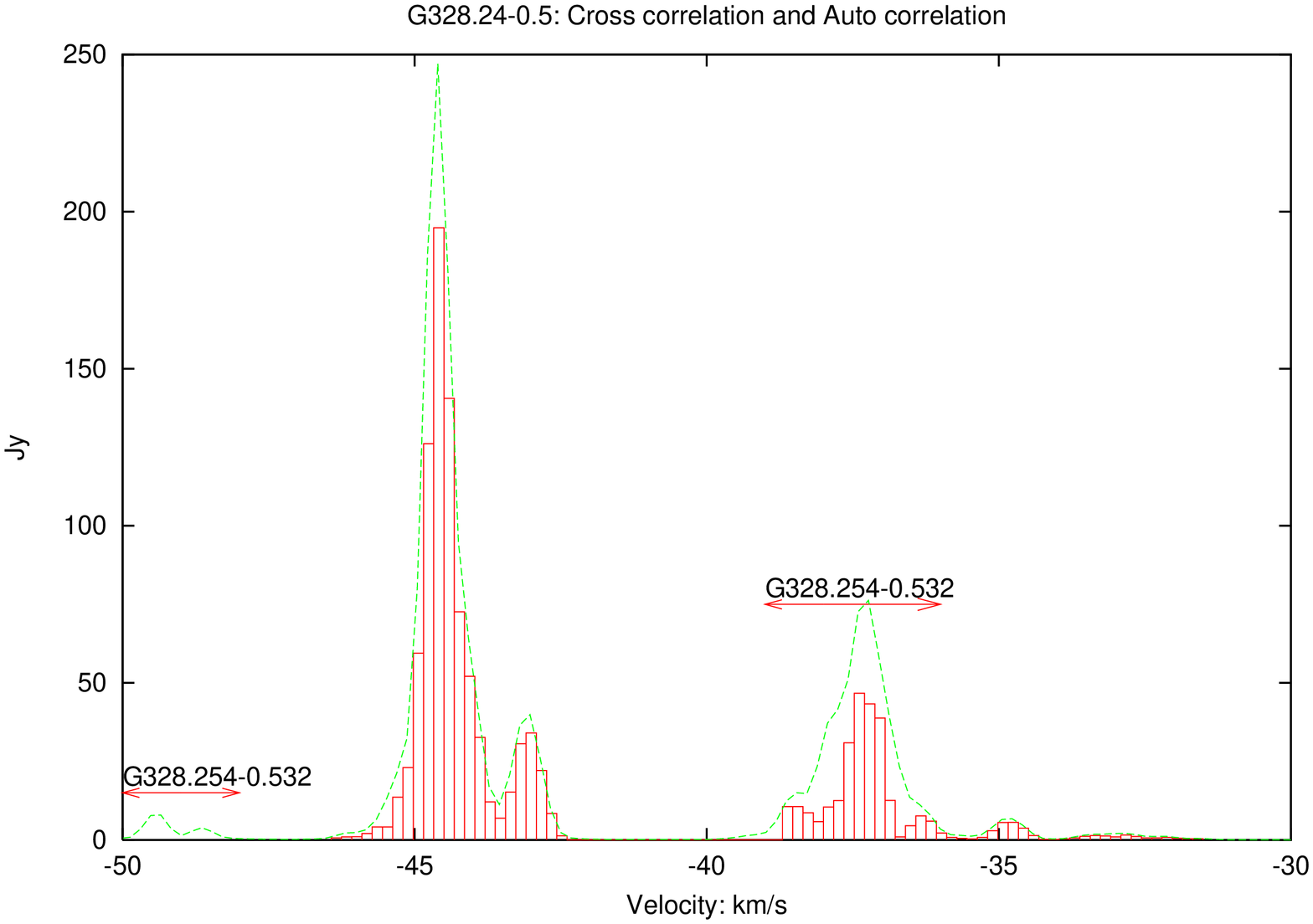, width=6cm, angle=270}
\caption{Autocorrelation (from Parkes with Antenna temperature
subtracted) (dotted line) against equivalent zero spacing flux density
(bar graph). This demonstrates the good flux recovery in the model
fitting.}
\label{fig:g328.24.ac}
\end{center}
\end{figure}


\begin{figure}
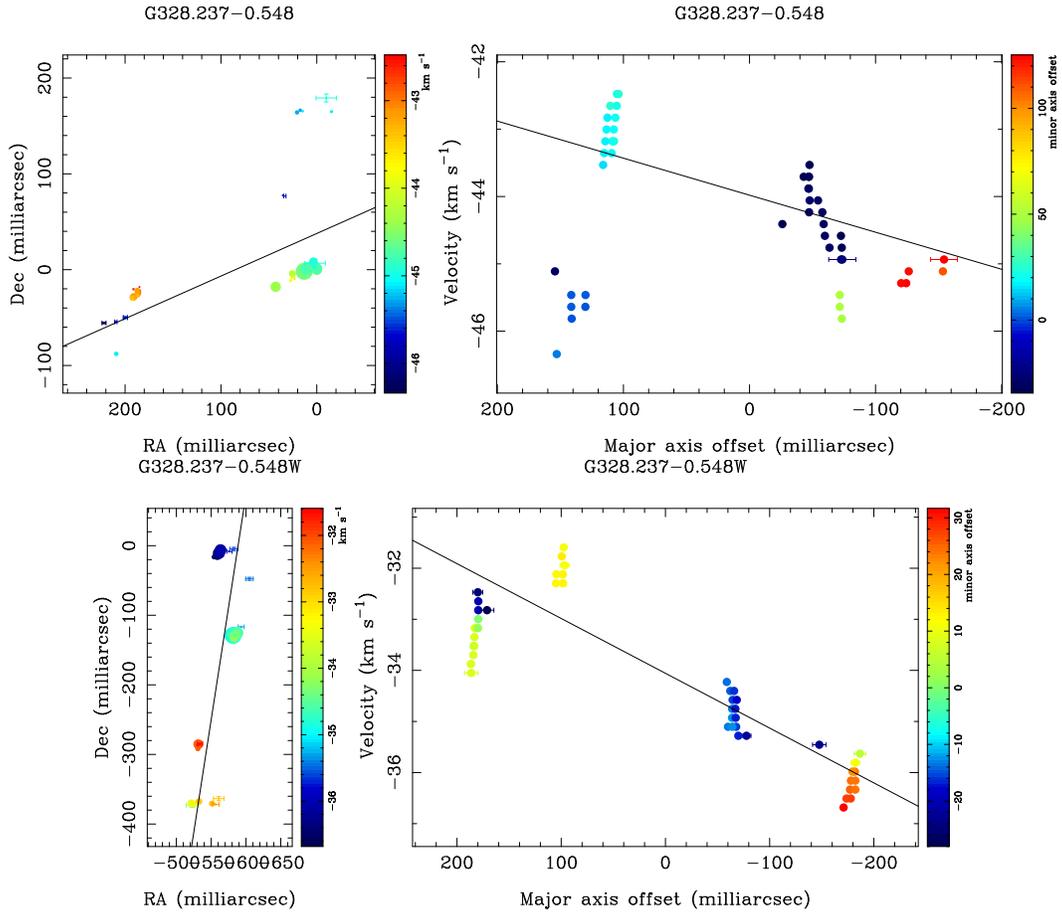

\begin{center}
\epsfig{file=g328.237-0.548.vlbi.ps, width=6cm, angle=270}
\epsfig{file=g328.237-0.548.va.ps, width=6cm, angle=270}
\epsfig{file=g328.237-0.548W.vlbi.ps, width=6cm, angle=270}
\epsfig{file=g328.237-0.548W.va.ps, width=6cm, angle=270}
\caption{a) Maser positions for clusters G328.237$-$0.548, b)
Velocity-Major Axis plot of the same data, c) Maser positions for clusters
G328.237$-$0.548W and d) Velocity-Major Axis plot of the same data. These follow the same conventions as Figure~1.}
\label{fig:g328.236}
\end{center}
\end{figure}



\begin{figure}
\begin{center}
\epsfig{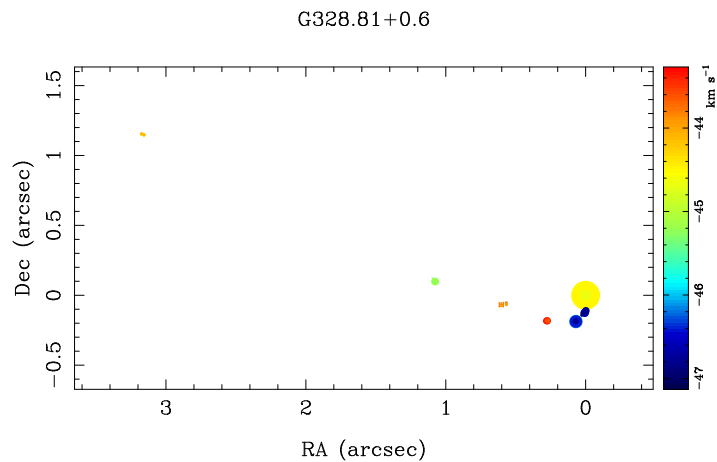}
\caption{Maser positions for clusters G328.809+0.633 and
  G328.808+0.633.  These follow the same conventions as Figure~1a.}
\label{fig:g328.81}
\end{center}
\end{figure}

\begin{figure}
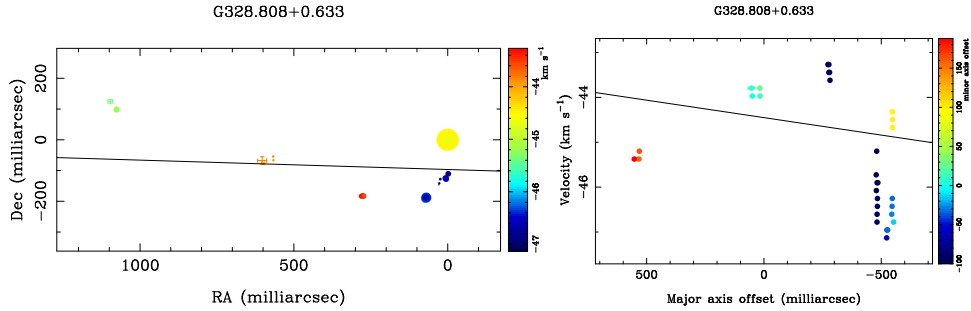

\begin{center}
\epsfig{file=g328.808+0.633.vlbi.ps, width=4cm, angle=270}
\epsfig{file=g328.808+0.633.va.ps, width=4cm, angle=270}
\caption{a) Maser positions for cluster G328.808+0.633 and b) Velocity-Major Axis
plot of the same data.  These follow the same conventions as Figure~1.}
\label{fig:g328.808}
\end{center}
\end{figure}

\begin{figure}
\begin{center}
\epsfig{file=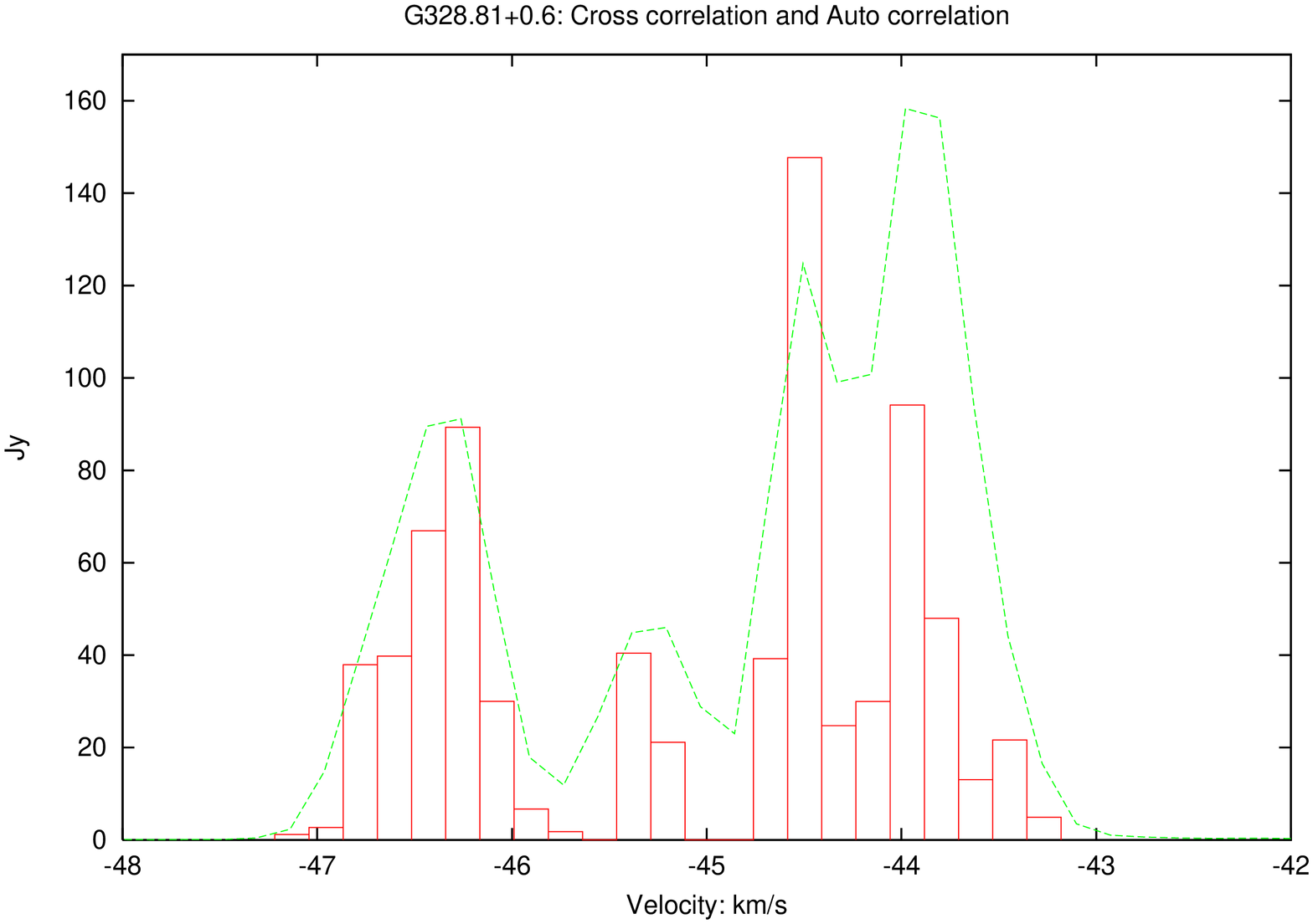, width=6cm, angle=270}
\caption{Autocorrelation (from Parkes with Antenna temperature
subtracted) (dotted line) against equivalent zero spacing flux density
(bar graph). This demonstrates
the good flux recovery in the model fitting.}
\label{fig:g328.81.ac}
\end{center}
\end{figure}

\begin{figure}
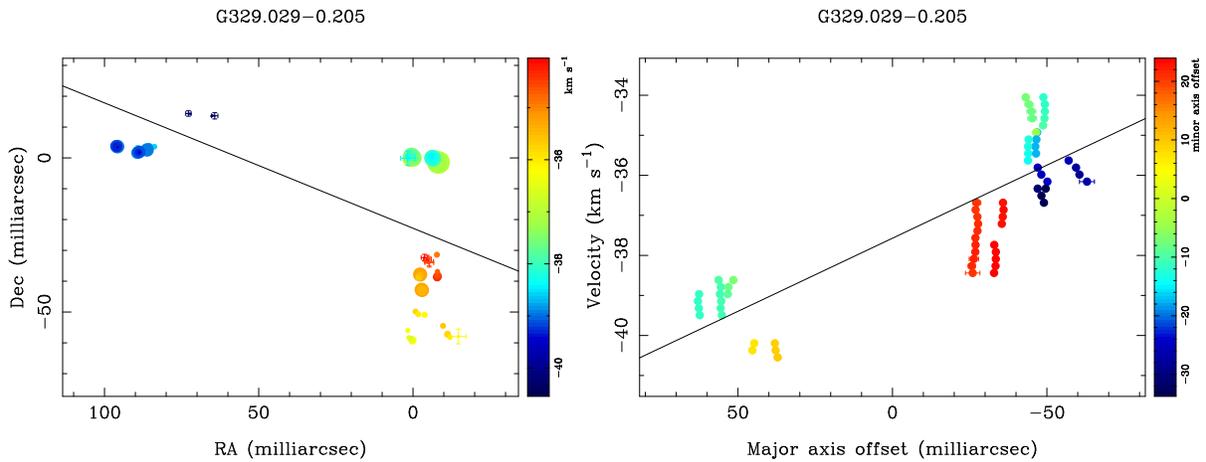

\begin{center}
\epsfig{file=g329.029-0.205.vlbi.ps, width=6cm, angle=270}
\epsfig{file=g329.029-0.205.va.ps, width=6cm, angle=270}
\caption{a) Maser positions for cluster G329.293-0.205 and b)
Velocity-Major Axis plot of the same data.  These follow the same
conventions as Figure~1.}
\label{fig:g329.03}
\end{center}
\end{figure}

\begin{figure}
\begin{center}
\epsfig{file=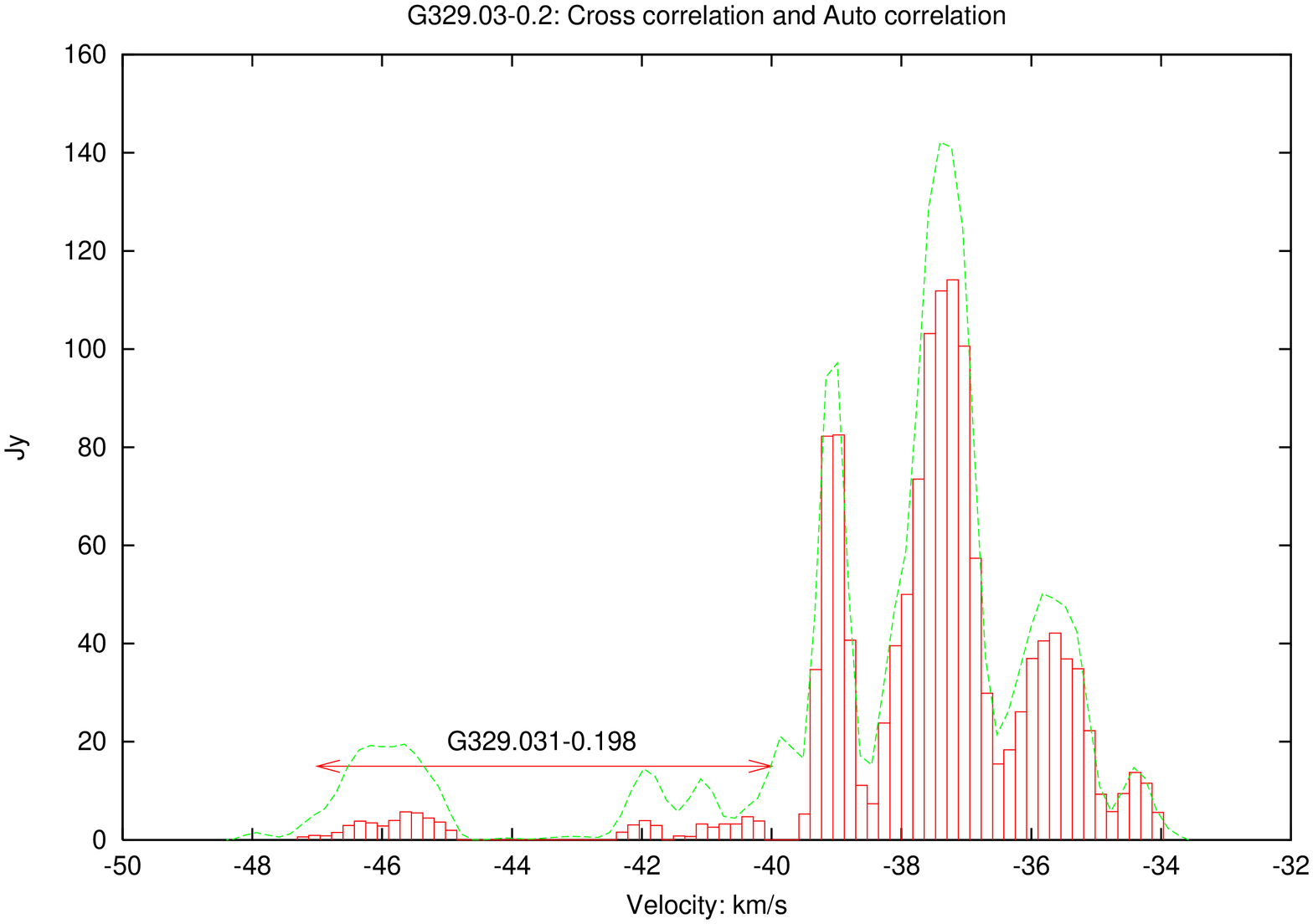, width=6cm, angle=270}
\caption{Autocorrelation (from Parkes with Antenna temperature
subtracted) (dotted line) against equivalent zero spacing flux density
(bar graph). This demonstrates the good flux recovery in the model
fitting.}
\label{fig:g329.03.ac}
\end{center}
\end{figure}

\begin{figure}
\begin{center}
\epsfig{file=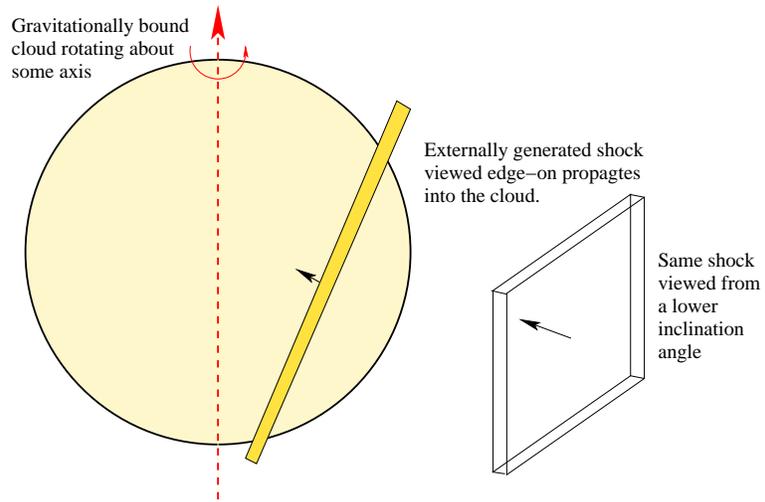, width=10cm}
\caption{Model of the shock and a self-gravitationally bound
  starforming region.}
\label{fig:model}
\end{center}
\end{figure}

\begin{figure}
\begin{center}
\epsfig{file=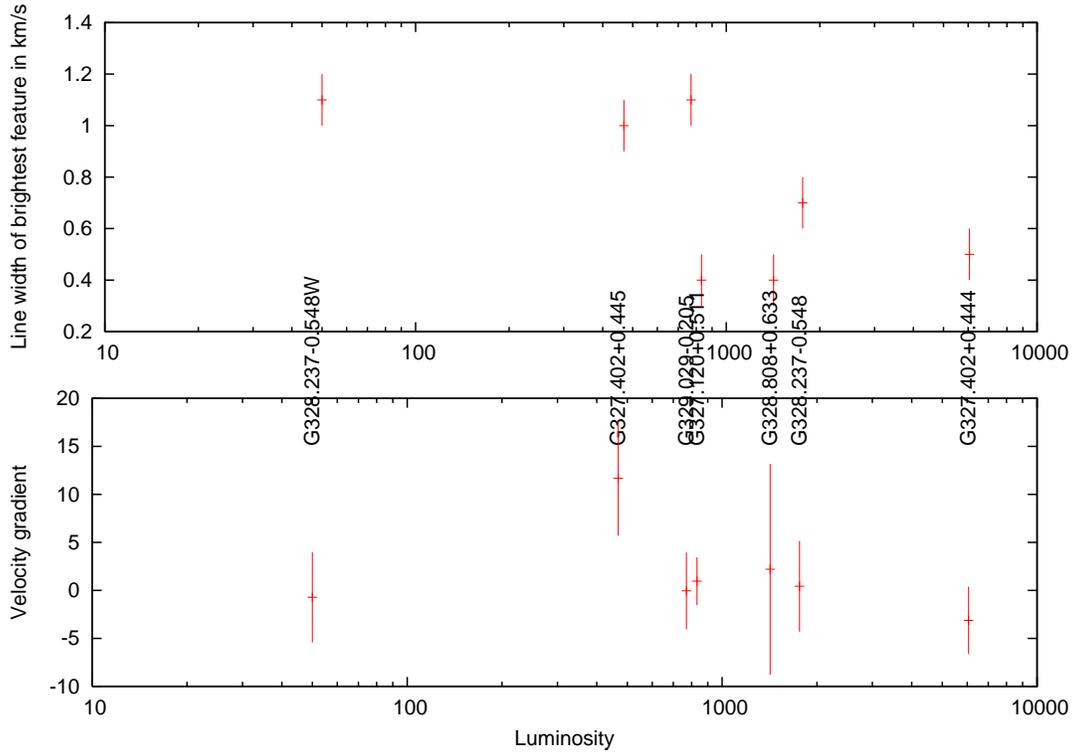,width=10cm,angle=270}
\caption{The comparison on the model predictions with the data. a)
  Peak emission line width and b) Median velocity gradients (in
  mas/\kms ) across the sources as a function of luminosity. No
  estimates of the errors in luminocity have been made, the errors in
  the linewidths are half the channel width and the errors in the
  gradients are the RMS of the individual gradients within each cluster.}
\label{fig:modfit}
\end{center}
\end{figure}


\end{document}